\documentclass[pdflatex,sn-mathphys]{sn-jnl}
\usepackage{caption}
\usepackage{subcaption}

\usepackage{amsmath}
\newcommand\bra[2][]{#1\langle {#2} #1\rvert}
\newcommand\ket[2][]{#1\lvert {#2} #1\rangle}

\usepackage{mathtools,array}

\usepackage{amsmath}

\jyear{2023}%

\theoremstyle{thmstyleone}%
\theoremstyle{thmstyletwo}%

\theoremstyle{thmstylethree}%

\definecolor{darkgreen}{rgb}{0,0.5,0}
\definecolor{tinygray}{rgb}{0.94,0.94,0.94}

\begin{document}

\title[Deep Quantum Circuit Simulation of Low-energy Nuclear States]{Deep Quantum Circuit Simulations of Low-energy Nuclear States}


\author[1,2]{\fnm{Ang} \sur{Li}}\email{ang.li@pnnl.gov}

\author[1]{\fnm{Alessandro} \sur{Baroni}}\email{baronia@ornl.gov}

\author[3]{\fnm{Ionel} \sur{Stetcu}}\email{stetcu@lanl.gov}

\author*[1]{\fnm{Travis S.} \sur{Humble}}\email{humblets@ornl.gov}

\affil[1]{\orgdiv{Quantum Science Center}, \orgname{Oak Ridge National Laboratory}, \orgaddress{\street{One Bethel Valley Road}, \city{Oak Ridge}, \postcode{37831}, \state{Tennessee}, \country{USA}}}

\affil[2]{\orgdiv{Physical and Computational Sciences Directorate}, \orgname{Pacific Northwest National Laboratory}, \orgaddress{\street{902 Battelle Blvd}, \city{Richland}, \postcode{99354}, \state{Washington}, \country{USA}}}

\affil[3]{\orgdiv{Theoretical Division}, \orgname{Los Alamos National Laboratory}, \orgaddress{P.O. Box 1663, MS B283, \city{Los Alamos}, \postcode{87545}, \state{New Mexico}, \country{USA}}}


\abstract{Numerical simulation is an important method for verifying the quantum circuits used to simulate low-energy nuclear states. However, real-world applications of quantum computing for nuclear theory often generate deep quantum circuits that place demanding memory and processing requirements on conventional simulation methods. Here, we present advances in high-performance numerical simulations of deep quantum circuits to efficiently verify the accuracy of low-energy nuclear physics applications. Our approach employs several novel methods for accelerating the numerical simulation including 1- and 2-qubit gate fusion techniques as well as management of simulated mid-circuit measurements to verify state preparation circuits. We test these methods across a variety of high-performance computing systems and our results show that circuits up to 21 qubits and more than 115,000,000 gates can be efficiently simulated. 
}



\maketitle


\section{Introduction}
\label{sec:introduction}


Quantum computing offers many opportunities to explore the complex interactions of many-body nuclear physics \cite{cloet2019opportunities,beck2023quantum}. This includes enabling efficient methods for modeling quantum physical processes as well as new techniques to simulate their outcomes \cite{Zhang_2021}. Recent discoveries in preparing quantum states to model nuclear physics have shown the efficacy of these methods for studying both structure and dynamics \cite{PhysRevC.104.024305,PhysRevResearch.4.043193,PhysRevC.105.064308,PhysRevC.105.064317}. In the low-energy regime, this includes calculating binding energies \cite{PhysRevLett.120.210501}, quadrupole moments \cite{PhysRevC.105.064318}, and electromagnetic transition rates \cite{PhysRevD.106.054508}. With early validation at small scales, quantum simulation methods for low-energy nuclear physics are expected to solve larger systems that are currently inaccessible to conventional approaches and thus expand computing capabilities for scientific discovery \cite{osti_1890248}
\par 
A powerful approach to these forms of digital quantum simulation rely on quantum circuits as sequences of gates to prepare and transform a quantum state under a defined Hamiltonian \cite{tacchino2020}. Accurate implementations of these quantum circuits are used to estimate observable outcomes that describe the prepared state \cite{cerezo2021variational}. The structure of the quantum circuits for quantum simulation are often derived from first-principles models of the underlying many-body problem that are then translated into exact or approximate operations \cite{Raeisi_2012}. For example, unitary coupled cluster (UCC) theory has been used recently to solve nuclear shell models by constructing quantum circuits that first encode fermionic fields in a spin representation and then apply unitary transformations to approximate the ground state \cite{PhysRevC.105.064308,pérezobiol2023nuclear}. By transforming these physics-driven quantum circuits into the spin representation, the subsequent unitary operator decomposition can be executed using a quantum computer \cite{Stetcu-2022proj,Kiss2023importancesampling}. Quantum circuits, therefore, represent the functionality of software programs that run on quantum computing hardware to perform these types of nuclear physics simulations.
\par 
An important step in designing the quantum simulation circuits is verification, which checks the correctness of the quantum circuit construction and the outputs observed during execution. 
Numerical simulation is a direct approach to verify the circuit transformations that can be used to check the prepared quantum state as well as the calculation of statistics and diagnostics \cite{wright2022numerical}. 
However, numerical simulation of quantum circuits is challenged by the exponential memory requirements with respect to system size for representing quantum states. Additionally, the probabilistic nature of quantum circuit execution often generates a combinatorial expansion of possible circuit outcomes that must be tracked. Formally, computational complexity arguments suggest that conventional numerical methods are intractable for simulating arbitrary quantum circuits, and it is widely anticipated that the applicability of numerical simulation is upper bounded by circuit width and depth, which is a concern for verification of quantum circuits. The continued development of numerical simulation techniques to accelerate and optimize such calculations is an area of active research within the quantum computing community.
\par 
The challenges for numerical simulation extend immediately to applications in nuclear physics, where deep quantum circuits spanning many qubits are often required to create highly accurate quantum states of the many-body problem. The case of deep quantum circuits is challenging for numerical simulation because these circuits are constructed from long sequences of unitary transformation, i.e., gates, to ensure accurate preparation of the intended quantum state \cite{PhysRevC.105.064308,pérezobiol2023nuclear}. Recent advances in using mid-circuit measurements as part of more efficient state preparation methods has helped to reduce circuit depth \cite{Stetcu-2022proj}, but this approach requires numerical simulation methods to manage the probabilistic outcomes generated by such measurements.
\par 
Here, we show how numerical simulations of deep quantum circuits can be used to efficiently verify the accuracy of low-energy nuclear physics applications. Our results demonstrate the performance scaling of quantum circuit simulations using novel methods to accommodate mid-circuit measurement integrated into deep circuit constructions. A principal approach is to leverage the right integration of GPU-accelerated numerical simulation methods. The leading contributions of this work are the development of mid-circuit measurement techniques to amplify the probability of observable outcomes; the improved efficiency of simulating  mid-circuit measurement by avoiding repeated circuit sampling, and the improved efficiency of deep circuit numerical simulations using gate fusion.
\par
The remainder is organized as follows. Section~\ref{sec:background} introduces background on the theoretical model for the second quantized Hamiltonian, the quantum algorithms for simulating the corresponding states, and methods and notation for numerical simulation of these quantum circuits. Section~\ref{sec:algorithm} specifies the algorithm for state preparation of low-energy nuclear states. Details about the numerical simulation methods are provided in Section~\ref{sec:simulator} with results presented in Section~\ref{sec:evaluation}. Concluding remarks are offered in Section~\ref{sec:conclusion}. 

\section{Background}
\label{sec:background}

\subsection{Theoretical Model and Qubit Mapping}

The nucleons inside the atomic nucleus interact via two-, three-, and so on body forces, with a clear hierarchy, in which two-body interactions dominate, followed by the many-body forces whose contributions diminish with increasing the number of bodies involved. 
A natural way to take into account the above constraint is given by chiral effective field theory ($\chi$EFT). 
In $\chi$EFT the symmetries of the underlying theory of strong interactions of quarks and gluons QCD are used to write down Lagrangians that describe interactions between nucleons and nucleons and pions and each interaction term is multiplied by a low energy constant that is fixed using some of the available experimental data. Interactions developed in this framework successfully predict nuclear binding energies up to $A=48$ \cite{Piarulli2018,Lonardoni2018,Maris2022}. 

In second quantization, the Hamiltonian governing the static and dynamics properties of the nuclear system can be written as

\begin{equation}
    H=\sum_{i,j=1,N_s} t_{ij} a^\dagger_i a_j +\frac{1}{2} \sum_{ij,kl} V_{ij,kl} a_i^\dagger a_j^\dagger a_l a_k+\cdots,
    \label{eq:hamSM}
\end{equation}
where $N_s$ is the number of single particle states included in the Hilbert space, $t_{ij}$ and $V_{ij;kl}$ are the one- and two-body matrix elements of the nucleon-nucleon interaction, while $a^\dagger_i$ ($a_i$) are the creation (annihilation) operators for state $i$. Given that the mean field is a good first approximation, we have allowed a general one-body term in Eq.~(\ref{eq:hamSM}). 

Using realistic full-space inter-nucleon interactions, like the ones rigorously constructed through $\chi$EFT, would not be feasible in the near future due to hardware limitations. Hence, a less demanding testing ground for quantum algorithms has  been identified in simpler models like the Lipkin-Meshkov-Glick model~\cite{PhysRevC.105.064317} or the phenomenological shell model \cite{PhysRevC.105.064308,pérezobiol2023nuclear}. In particular the shell model offers the advantage that it has some of the complexity of realistic nucleon-nucleon interactions in a small model space, even though it was found to produce very entangled states \cite{PhysRevC.105.064308}. The simulations presented in this paper were performed using the Cohen-Kurath interaction \cite{COHEN19651} in the $p$ shell, where active the space is restricted to six proton and neutron states with $j=1/2$ and $j=3/2$, assuming a $^4$He inert core. We would like to emphasize that while an excellent testing ground for quantum algorithms and even hardware benchmarks, shell model calculations in larger model spaces is not our final goal but rather an intermediary step for the time being.

The first step in the process of implementing various algorithms on quantum hardware and/or numerical simulators is to map this interaction to the qubit space, and different mappings could have advantages over the others, although the advantage is not always clear nor easy to quantify. The Jordan-Wigner (JW) \cite{JW} or occupation mapping is the most intuitive second quantization scheme. In this case, the creation and annihilation operators are given by 
\begin{eqnarray}
a^\dagger_i = && \frac{1}{2} \left(\prod_{j=0}^{i-1}Z_j\right) (X_i-iY_i),\label{eq:mapJWd}\\
a_i = && \frac{1}{2}\left(\prod_{j=0}^{i-1}Z_j \right)(X_i+iY_i), \label{eq:mapJW}
\end{eqnarray}
where the $X_i$, $Y_i$ and $Z_i$ are the Pauli matrices acting on qubit $i$. Thus, each single-particle state is associated with a single qubit, and in order to describe a system with $N_s$, the number of qubits required is $N_s$. If the qubit is measured in state $\ket 0$, the state is unoccupied, while a measurement in state $\ket 1$ represents an occupied state.

The Jordan-Wigner mapping is very simple and intuitive, but it has a couple of disadvantages. First, it describes a Fock space of dimension $2^{N_s}$ that contains particle numbers from zero to $N_s$. For  particle-conserving Hamiltonians as the ones governing the nuclear many-body system, only one subspace with a fixed particle number is active at one time. Moreover, for the nuclear problem, the JW mapping is even less efficient given that the Hamiltonian concurrently preserves the proton and neutron particle numbers. Let us take an example considering $N_p$ protons and $N_n$ neutrons, each in $N_s$ states. In this case, for large $N_s$, and using the Stirling approximation \cite{abramowitz+stegun}, the number of many-body states with $N_p$ protons and $N_p$ is
\begin{eqnarray}
 \lefteqn{\tilde N \approx \left( \frac{N_s^2}{(N_s-N_p)(N_s-N_n)} \right)^{N_s} } \nonumber \\
  &&  \times (N_s-N_p)^{N_p}(N_s-N_n)^{N_n}e^{-N_p-N_n},
  \label{eq:Nstates_part}
\end{eqnarray}
which is much smaller than the dimension of the Fock space $2^{2N_s}$ represented in the Jordan Wigner mapping. Second, in order to represent state $i$ one must include $i-1$ operators acting on the previous qubits, and hence, the number of one-qubit gates and entanglement gates in general is quite large. This poses challenges for applications on current and near-term devices.

While less transparent, other more efficient mappings within the second quantization framework exist. One alternative is the Bravyi-Kitaev mapping \cite{BRAVYI2002210}, where the qubits store partial sums of occupation numbers, requiring the number of states to be powers of 2. This mapping can be as efficient, and in many cases even more efficient in quantum chemistry calculations of ground states of molecular systems \cite{Tranter:2015,Tranter:2018}. In the parity representation, the $n^\mathrm{th}$ qubit stores the sum of the parity of the first $n$ modes \cite{Seeley:2012,RevModPhys.92.015003}. However, in applications to quantum chemistry, this scheme was not found to be particularly useful \cite{RevModPhys.92.015003}.

A first quantization approach would be more efficient in terms of the number of qubits required to describe the system. In this case, any particular one single particle state is encoded as a fixed combination of many qubit states, with the number of required qubits being given by $\approx \log_2(N_s)$. In each case, to describe each particle one needs the same number of qubits, so that the total number of qubits would be $N_q\approx n\log_2(N_s)$, with $n$ the number of particles. It is clear that for this mapping, the scaling is more advantageous as $n\log_2(N_s)\ll N_s$ for $n\ll N_s$. The main disadvantage of this mapping, however, is that the many-body system is not automatically antisymmetric for a system of fermions.  For a large number of particles this is especially challenging. 
However, a recent quantum algorithm~\cite{Berry2018a} was proposed for antisymmetrizing identical fermions with a gate complexity of $O(\log^cn \log_2\log_2 N_s)$ and a circuit size of $O(n\log^cn\log_2 N_s)$. The value of $c$ depends on the choice of the specific algorithm.

The advantages of each mapping scheme are not obvious. One possible middle ground between the second and the first quantization mapping could be the scheme proposed in Ref.~\cite{PhysRevResearch.4.023154}, in which one preserves the antisymmetrization of the basis states, thus eliminating the need for a complicated antisymmetrization first step. In addition, while the main justification for the work in Ref.~\cite{PhysRevResearch.4.023154} is particle-number conservation, other symmetries can be considered as well. Thus, this approach is perfectly fitted for the shell model. In an occupation basis, the Hamiltonian in Eq.~(\ref{eq:hamSM}) becomes
\begin{equation}
    H=\sum_{\alpha,\beta}\bra \alpha H\ket\beta \ket\alpha \bra\beta,
\end{equation}
where $\ket\alpha$, $\ket\beta$ are many-body states that preserve not only the particle number, but also the projection of the total angular momentum in the $M$ scheme. For the $j-j$ coupling, the many-body states $\ket\alpha$ and $\ket\beta$ have good total angular momentum; in this case, the matrix $H$ is block diagonal, as $[H,J^2]=0$. In the $j-j$ scheme the matrix that needs to be diagonalized is less sparse than in $M$ scheme, but the dimension is smaller. While it is unclear how the trade off between the dimension that needs to be solved would map into quantum hardware, it is clear that many of the tools \cite{MFDcode,BROWN2014115,LSU3shell,BIGSTICKSM:2018} developed for large-scale shell model calculations can be adopted for porting the same type of problems on quantum hardware. 

Basis states generated in $M$ or $j-j$ schemes are assigned to a certain combination of qubits. A similar mapping was used in Ref. \cite{PhysRevA.103.042405} to solve for the deuteron in relative coordinates in a harmonic oscillator basis. In Table \ref{tab:mapping}, we compare the number of qubits necessary to represent a nuclear system with $N_p$ protons and $N_n$ neutrons for Jordan-Wigner mapping ($N_q^\mathrm{JW}$) and using encoding of the basis states into combinations of qubits ($N_q^\mathrm{SM}$), with a fixed projection of the spin $M_{tot}$. It is immediately clear that while it might be more advantageous to encode the shell model basis, the number of Pauli strings in this mapping ($N_\mathrm{Pauli}^\mathrm{SM}$) quickly surpasses the number of Pauli strings necessary for the JW mapping ($N_\mathrm{Pauli}^\mathrm{JW}$). Thus, the advantage of using the qubit-efficient enconding quickly goes away in particular for the $sd$ shell model space, where the number of single-particle states included in the calculation increases to 12 for each species, assuming an inter $^{16}$O core, and using the ``universal" SD interaction \cite{WILDENTHAL19845,PhysRevC.74.034315}. Note that in Table \ref{tab:mapping} we used an arbitrary order for encoding, and not the one based on the Gray code that could potentially be more efficient~\cite{PhysRevA.103.042405}. Nevertheless, we do not expect the picture presented in Table \ref{tab:mapping} to dramatically change.

\begin{table}[!t]
\centering\small
\caption{Comparison interaction mapping and number of qubits for several $p$ and $sd$ shell nuclei. $M_{tot}=0$ for an even number of particles, $M_{tot}=1/2$ for an odd number of particles.}
\begin{tabular}{ccccccc}
\hline\hline
Shell & $N_p$ & $N_n$ & $N_q^\mathrm{JW}$ & $N_\mathrm{Pauli}^\mathrm{JW}$ & $N_q^\mathrm{SM}$ & $N_\mathrm{Pauli}^\mathrm{SM}$  \\ \hline
$p$ & 1 & 2 & 12 & 975 & 5 & 528 \\
$p$ & 2 & 2 & 12 & 975 & 6 & 2,072 \\
$p$ & 1 & 3 & 12 & 975 & 5 & 488 \\
$p$ & 2 & 3 & 12 & 975 & 6 & 2,080 \\
$p$ & 3 & 3 & 12 & 975 & 7 & 7,936 \\
$sd$ & 1 & 2 & 24 & 12,869 & 7 & 8,252 \\
$sd$ & 1 & 3 & 24 & 12,869 & 9 & 131,321 \\
$sd$ & 2 & 2 & 24 & 12,869 & 10 & 523,720 \\

  \hline\hline
\end{tabular}
\label{tab:mapping}
\end{table}

\section{Algorithm Design}
\label{sec:algorithm}

\subsection{Projection Algorithm for State Preparation}
\label{subsec:proj}

Several algorithms are now available for preparing selected states on quantum hardware \cite{Ge2018,Motta:2020,Choi2021,Jouzdani2022,Stetcu-2022proj}. A projection algorithm like the one introduced in Ref.~\cite{Stetcu-2022proj} could be better suited to the nuclear physics problems, as it can be adapted to take into account even limited information about the nuclear spectrum. 

All projection algorithms use a series of time evolutions and measurements of one or more ancilla qubits connected to the system in order to project on a desired state. To project the ground state, in Ref.~\cite{Ge2018}, Ge et al. apply $\cos^M(\tilde H\pi/2)$ on a trial state, with $\tilde H$ shifted and re-scaled Hamitlonian $H$ and $M$ an integer. Thus, if $M$ is large enough, the amplitude of any arbitrary state with energy $E$ is reduced by $\cos^M(E\pi/2)$, ensuring the suppression of all but the state at $E=0$. In a very simplistic approach, applying this type of filtering reduces to time evolution with constant time $t=\pi/2$ and measuring $M$ ancilla states. A more efficient implementation based on $\log_2(M)$ ancilla qubits is presented in Ref.~\cite{Ge2018} and its implementation is discussed in Appendix \ref{app:cirac_algo}.  In contrast, the Rodeo algorithm \cite{Choi2021} exploits the fact that the probability of finding a state with energy $E$ given by 
\begin{equation}
P_N=\prod_{n=1}^N\cos^2\left[(E_\mathrm{target}-E)\frac{t_n}{2}\right]
\end{equation}
is suppressed by a factor $1/4^N$ if $E\neq E_\mathrm{target}$ for a Gaussian distribution of times $t_n$ and large $N$, with $N$ the number of measurements. In the proposed implementation, the Rodeo algorithm is based on $N$ controlled time evolution with the Hamiltonian governing the dynamics and normal distributed times. 


Neither of the two algorithms briefly presented above uses any information about the system one wants to describe. It is well understood in imaginary time evolution approaches in classical calculations that in order to filter out contributions from undesired states, the evolution time is of the order of $1/\Delta$, where $\Delta$ is the gap between the target state and the first state with non-zero overlap with the trial state. A direct application on quantum computers of the imaginary time evolution is cumbersome, but the same general lessons can be applied using real-time evolution. In Ref.~\cite{Stetcu-2022proj}, we introduced a projection algorithm based on real-time evolution and using a set of optimized times and small phases. The optimization is based on the approximate knowledge of the many-body spectrum and the general assumption regarding the overlap of the eigenstates with the trial state. Since this approach was much more efficient in state preparation than other projection-based algorithms \cite{Stetcu-2022proj} even for small gaps, in this work we use this procedure to construct the ground state of a many-body system.

In the simplest implementation of the energy filtering algorithm of Ref.~\cite{Stetcu-2022proj}, a single ancilla qubit is attached to the collection of qubits describing the physical system. The trial state $\ket {\psi_0}$ is usually a Hartree-Fock Slater determinant, which can be trivially represented on a quantum computer, and the ancilla qubit $a$ is set to state $\ket{0}$. Then, one performs a series of time evolutions with times $t_i$ followed by a measurement of the qubit $a$. The algorithm is successful if all measurements of the ancilla qubit produce state $\ket{0}$. At step $i$ one produces the state $\ket{\psi_i}$ from state $\ket{\psi_{i-1}}$ as follows
\begin{eqnarray}
    \lefteqn{
    \ket{\psi_i}=\exp\left[-i(\tilde H t_i +\delta_i) Y_a\right]\ket{\psi_{i-1}}\otimes\ket{0} }\nonumber\\
    & & = \cos{\left(\tilde H t_i +\delta_i \right)}\ket{\psi_{i-1}}\otimes\ket{0}+
    \sin{\left(\tilde H t_i +\delta_i \right)}\ket{\psi_{i-1}}\otimes\ket{1}.
    \label{eq:proj}
\end{eqnarray}
Knowledge of the spectrum, even approximative, is not required, but is useful. For example, the same algorithm can be used to project on quantum numbers. While projection on symmetries has been introduced before \cite{PhysRevLett.125.230502,PhysRevC.105.024324}, the algorithm based on Eq.~(\ref{eq:proj}) could be more efficient as a large number of quantum numbers can be eliminated at each iteration \cite{Stetcu-2022proj}. Moreover, since the projection on targeted quantum numbers usually increases the gap that controls the algorithm, the projection can become more effective with smaller propagation times, akin to classical calculations.

In the investigations presented in this paper, we did not perform symmetry projection. In turn, we have only projected the ground state, assumed at zero energy. To better understand the algorithm in Eq.~(\ref{eq:proj}), and the optimization procedure, we start with the trial vector decomposed in an orthogonal basis 
\begin{equation}
\ket{\psi_0}=\sum_\alpha C_\alpha\ket \alpha,     
\end{equation}
where $\ket \alpha$ are the (unknown) eigenstates of $\tilde H$ ($\tilde H\ket \alpha=E_\alpha \ket \alpha$), and $C_\alpha$ (unknown) complex coefficients.  Time evolving the system using Eq.~(\ref{eq:proj}) produces the state
\begin{eqnarray}
    \lefteqn{
    \ket{\psi_1}=\exp\left[-i(\tilde H t_1 +\delta_1) Y_a\right]\ket{\psi_{0}}\otimes\ket{0} }\nonumber\\
    & & = \sum_\alpha C_\alpha\ket{\alpha}\otimes\left[\cos{\left(E_\alpha t_1 +\delta_1 \right)}\ket{0}+
    \sin{\left(E_\alpha t_1 +\delta_1 \right)}\ket{1}\right].
    \label{eq:proj1}
\end{eqnarray}
Since the ground state is shifted so that $E_0=0$, if the phases $\delta_i$ are kept small, measuring the ancilla qubit in state $\ket 0$ will enhance the amplitude of the ground state with respect to the other states. If the gap is known from classical calculations, one can always take the time $t_1=\pi/(2\Delta)$ and $\delta_1=0$, so that after the first measurement, the state at the gap is exactly removed from physical state (without the ancilla qubit). In the case the spectrum is known, one can continue like in the case of projection on quantum numbers. However, for the more general case when the spectrum is not known, it was found that additional exponentially shorter times ($t_i=t_{i-1}/2$) are a reasonable choice to remove higher lying excitations \cite{Stetcu-2022proj}, with a total evolution time approaching $2t_1$. If the suppression of the unwanted states is deemed unsatisfactory, one can repeat the same procedure, while running the circuit in Fig. 12 of Ref.~\cite{Soma-SpectrumState} will provide  information about the gap and other states present in the trial state, if desired. 

The probability to produce the desired state is given by the probability to find that state into the trial state in the case of all phases zero. Using Eq.~(\ref{eq:proj1}), the amplitude of each state after $N$ measurements of the ancilla (all giving state $\ket 0$) becomes $C_\alpha'=\prod_{i=1}^N\cos(E_\alpha t_i+\delta_i)C_\alpha$. To optimize the times and phases in order to speed up the calculation, one can start with the exponentially distributed times, assuming some reasonable overlap with the ground state, and random overlaps for the remaining states, and then maximize the final overlap for the final state. This optimization procedure can be classically performed, and general properties regarding the nuclear spectra can be used in the case when the spectrum is not \textit{a priori} known. If non-zero phases are included in the optimization, they are generally small, and reduce the probability of success for the targeted state by $\prod_{i=0}^N\cos^2(\delta_i)$.

\subsection{Numerical Simulation of Quantum Circuit}

\begin{table}[!t]
\centering\footnotesize
\caption{A summary of numerical methods for simulating quantum circuits categorized by memory cost for qubits (Mem(Q)), memory cost for gates (Mem(G)), computational cost for qubits (Comp(Q)), and computational cost for gates (Comp(G)). Sparsity identifies if sparsity in the representation can be utilized, while Noise identifies if noise can be simulated.}
\begin{tabular}{|c|c|c|c|c|c|c|c|}
\hline
\textbf{Approach} & \textbf{Mem(Q)} & \textbf{Mem(G)} & \textbf{Comp(Q)} & \textbf{Comp(G)} & \textbf{Sparsity} & \textbf{Noise} & \textbf{Ref}  \\ \hline
State Vector & High & Low & High & Low & No & No & \cite{li2021sv} \\ \hline
Density Matrix &  High & Low &  High & Low & No & Yes & \cite{li2020density} \\ \hline
Decision Diagram & Medium & Medium & Medium & High & Yes & No & \cite{grurl2021stochastic} \\ \hline
Tensor Network & Medium & High & Medium & High & Yes & No & \cite{nguyen2021tensor} \\ \hline
Stabilizer & Low & Low & Low & Low & Yes & No & \cite{aaronson2004improved} \\ \hline
Device Simulation & High & High & High & High & No & Yes & \cite{fu2017experimental} \\ \hline
\end{tabular}
\label{tab:sim_approach}
\end{table}

Generally speaking, there are multiple ways to numerically simulate the quantum circuits generated from a quantum algorithm, such as the projection algorithm described. These include state-vector \cite{jones2019quest, li2021sv}, density-matrix \cite{li2020density, chen2021low}, tensor-network \cite{markov2008simulating, nguyen2021tensor}, decision diagrams \cite{miller2006decision, grurl2020considering}, stabilizer \cite{aaronson2004improved, bravyi2019simulation}, and device-level simulation such as pulse-based simulation \cite{mckay2018qiskit}. Some of their features are summarized in Table~\ref{tab:sim_approach}, with respect to the cost of system memory and computation when scaling qubits ($Q$) and gates ($G$), as well as the ability to leverage sparsity in the representation and incorporating noise effect. 

As discussed, the circuit for the time evolution and projection algorithm can be quite deep which is far beyond the capability and coherence time of current NISQ devices. For algorithm verification purposes without inspecting noise effects, in this work we mainly focus on state-vector simulation given its resilience to circuit depth. In the following, we briefly introduce the fundamentals of state-vector representation and the general ways of performing state-vector based numerical simulation in a classical machine.

\vspace{4pt}\noindent\textbf{State-Vector Representation:} A pure quantum state in a mathematical formulation corresponds to a vector in the Hilbert space. A quantum system in a superposition state thus can be represented as a linear combination of the (orthonormal) eigenstates according to some basis: 
\begin{equation*}
\ket{\Psi(t)} = \sum_s{C_s(t)\ket{\Phi_s}}
\end{equation*}

The coefficient $C_s$ which corresponds to a particular eigenstate is a complex number thus allowing interference effects among states. Evolution of the quantum states is time dependent, governed by the evolution operators defined in the quantum circuit. State transitions are triggered by evolution operators of which each represents a gate. Through the gate sequence of the circuit, the quantum system evolves toward an objective state of interest, where measurement can be repeatedly applied for sampling the state. The squares of the coefficients sum-up to 1. 
The simulation approach for pure state is to use an array with complex numbers to represent the coefficients $C_s$, known as the state-vector. The size of the array depends on the number of eigenstates in the system, which scales as $2^n$ with respect to the number of qubits $n$. To ensure numerical accuracy, double-precision is necessary. The system is evolved by applying the gates, each denoting a unitary matrix that describes how the coefficients of certain eigenstates need to be adjusted. Once a gate is applied, the system transits to the next state:
\begin{equation}
  \ket{\Psi}\to U\ket{\Psi} 
\end{equation}

The state-vector approach describes the fundamental evolution of a quantum system in an ideal scenario without noise impact, thus is widely used for quantum algorithm verification, which is the purpose of this work.

\vspace{4pt}\noindent\textbf{State-Vector Simulation:} State-vector based simulation is to simulate the operations of applying a series of unitary operators $U_{m-1} \cdots U_1 U_0$ to the state-vector $\ket{\psi}=\sum_{i=0}^{2^n-1}\alpha_i\ket{i}$ that describes the state of a quantum system. $n$ is the number of qubits and $m$ is the number of operations or gates. Here, a complex-valued double-precision floating-point vector $\vec{\alpha}$ of size $2^n$ is used to store the coefficients $\alpha_i$, which costs $16\times2^n$ bytes of memory in a classical computer. $U_i$ with $i\in[0,m-1]$ is a $2\times2$ (for one-qubit gate) or $4\times4$ (for two-qubit gate) complex matrix. It has been shown that an arbitrary quantum circuit can be decomposed into 1-qubit and 2-qubit gates \cite{barenco1995elementary}. In fact, almost all real quantum devices execute 1-qubit or 2-qubit basis gates internally. For example, IBMQ adopts 1-qubit gate \texttt{X}, \texttt{SX}, \texttt{RZ}, \texttt{ID} and 2-qubit gate \texttt{CX} as the basis gates; Rigetti uses 1-qubit gate \texttt{RX}, \texttt{RZ} and 2-qubit gate \texttt{CZ} for internal execution. Multi-qubit gates are decomposed into 1-qubit and 2-qubit gates.

To apply a gate $U$, the operation is $\ket{\psi}\to U\ket{\psi}$. For 1-qubit $U$ applying on qubit $q$ in a quantum register, $\vec{\alpha}$ is updated through the following expression where $s_i=\lfloor i/{2^q}\rfloor 2^{q+1}+(i \% 2^q)$ for every integer $i\in[0, 2^{n-1}-1]$:
\begin{equation}
\begin{bmatrix}
\alpha_{s_i} \\
\alpha_{s_i+2^q}  
\end{bmatrix}
\to
U_{2\times2}\cdot \begin{bmatrix}
\alpha_{s_i} \\
\alpha_{s_i+2^q}  
\end{bmatrix}
\label{eq:1q-op}
\end{equation}
Regarding 2-qubit unitary gate $U$ applying on qubit $p$ and $q$ (assuming $p<q$ without losing generality), $\vec{\alpha}$ is updated  through:
\begin{equation}
\begin{bmatrix}
\alpha_{s_i} \\
\alpha_{s_i+2^p} \\
\alpha_{s_i+2^q} \\
\alpha_{s_i+2^p+2^q}
\end{bmatrix}
\to
U_{4\times4}\cdot \begin{bmatrix}
\alpha_{s_i} \\
\alpha_{s_i+2^p} \\
\alpha_{s_i+2^q} \\
\alpha_{s_i+2^p+2^q}
\end{bmatrix}
\label{eq:2q-op}
\end{equation}
where $s_i=\lfloor \lfloor i/{2^p}\rfloor /2^{q-p-1} \rfloor  2^{q+1} + (\lfloor i/{2^p}\rfloor \% 2^{q-p-1})2^{p+1} +  (i \% 2^p)$ for every integer $i\in[0,2^{n-2}-1]$. 

To summarize, state-vector based quantum numerical simulation is to perform a sequence of $2\times2$ or $4\times4$ operations Eq.~(\ref{eq:1q-op}) and Eq.~(\ref{eq:2q-op}) over the large state-vector coefficient array of complex numbers. Note, although quantum gates without noise are all unitary operations, Eq.~(\ref{eq:1q-op}) and Eq.~(\ref{eq:2q-op}) can be more general and do not necessarily require $U$ to be unitary, which offers the great opportunities of gate fusion, as will be discussed in Section~\ref{subsec:gate_fusion}.

\section{Numerical Simulator Design}
\label{sec:simulator}

Our state-vector numerical simulator for low-energy nuclear state-preparation is developed from our previous work SV-Sim~\cite{li2021sv} of the NWQSim package. Considering the deep circuits and the unique amplification based algorithm through mid-circuit ancilla measurement, in this work, we propose two techniques for efficient simulation. We first introduce the simulator framework and then focus on the two techniques.   

\subsection{Simulator Framework}

\begin{figure}[!t]
\centering
\includegraphics[width=0.75\columnwidth]{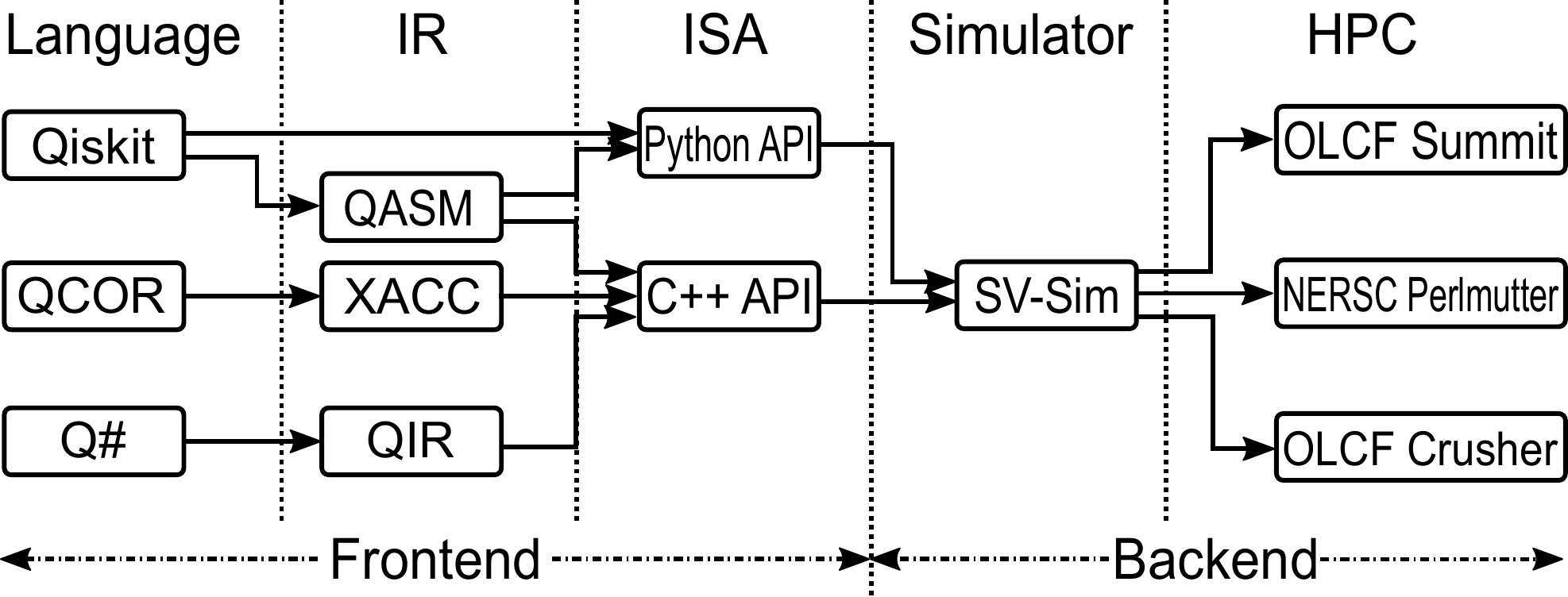} 
\caption{Simulation Framework.}
\label{fig:svsim_arch}
\end{figure}

Fig.~\ref{fig:svsim_arch} illustrates the SV-Sim framework~\cite{li2021sv}. It offers both C++ and Python interfaces to support quantum programming environments such as Qiskit~\cite{cross2018ibm},  Q\#~\cite{qsharp}, QCOR~\cite{mintz2020qcor}, as well as quantum intermediate representations (IR) such as QASM~\cite{cross2017open} and QIR~\cite{qir}. The backends include CPUs, GPUs, Xeon-Phis, and multi-node heterogeneous HPC clusters. 

In terms of single-device backend implementation, the original SV-Sim harvests performance of heterogeneous accelerators such as GPUs through two major strategies: 
\emph{homogeneous execution}, where GPU-side polymorphism is realized through a device functional pointer approach~\cite{li2021sv}, so that all the gate operations on the GPU side can be merged into a single GPU kernel, avoiding kernel creation, context switching, and data movement overhead. \emph{Gate-specialized implementation}, where the speciality of the gate matrices, including sparsity and identity, are exploited. The new simulator designed in this work still benefits from homogeneous execution, but we could not use gate-specialization anymore due to gate fusion, as will be seen in Section~\ref{subsec:gate_fusion}.

\begin{figure}[!t]
\centering
\includegraphics[width=0.7\columnwidth]{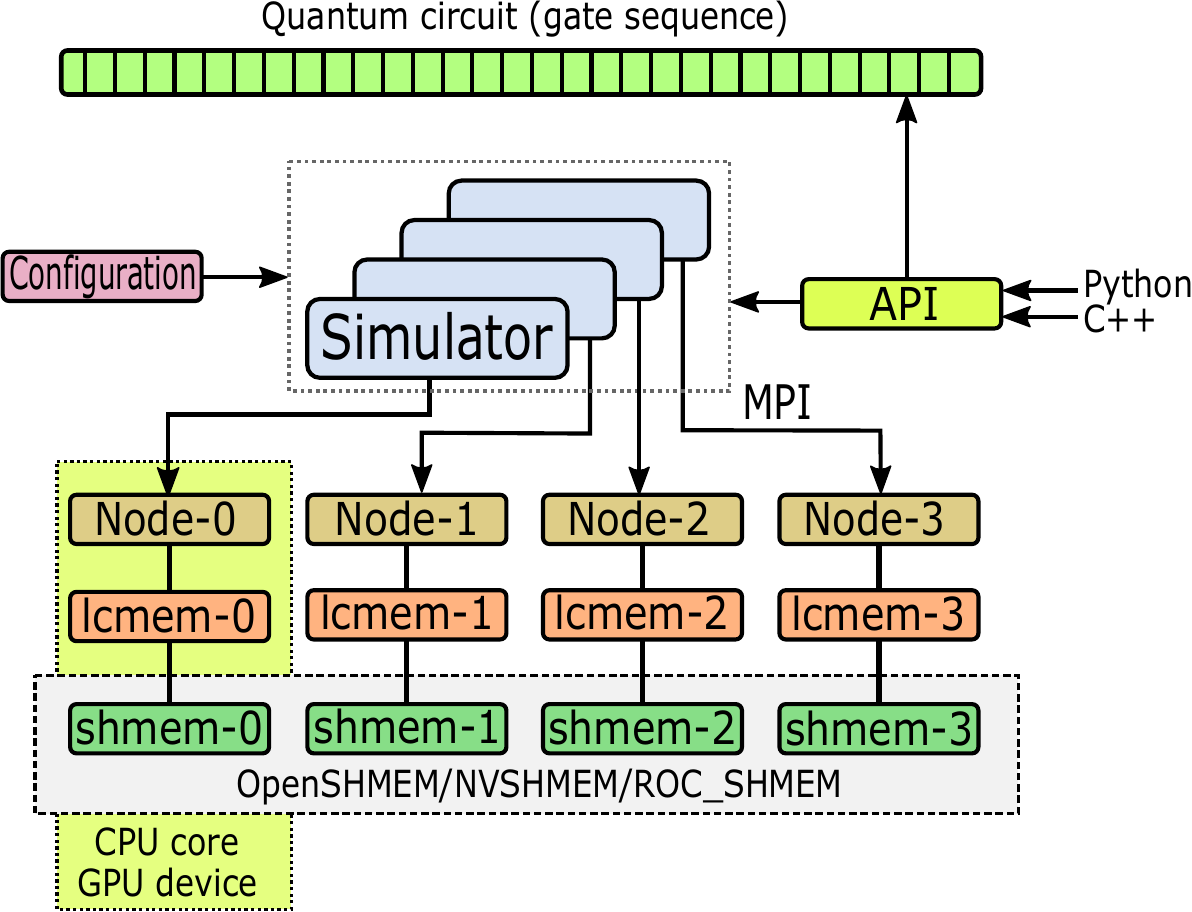} 
\caption{Architecture for the scaling-out of the SV-Sim simulator.}
\label{fig:inter-node}
\end{figure}

In terms of the cluster backend implementation, SV-Sim mainly performs the communication through the shared-memory (SHMEM) model and interfaces, such as NVSHMEM and OpenSHMEM. MPI is also used when necessary. Fig.~\ref{fig:inter-node} shows the architecture for the cluster implementation. Each SHMEM or MPI process owns a simulator object. Therefore, rather than using a single simulator instance to manage the multi-devices, each of them operates on a single CPU or GPU. The state-vector coefficients are evenly distributed among the devices. SHMEM/MPI are used for inter-device communication.

\begin{table}[!t]
\centering\footnotesize
\caption{Gates defined in IBM OpenQASM~\cite{cross2017open}.}
\begin{tabular}{|c|l|c|l|}
\hline
\textbf{Gates} & \textbf{Meaning} & \textbf{Gates} & \textbf{Meaning}  \\ \hline
\texttt{U3} & 3 parameter 2 pulse 1-qubit  &  \texttt{CY} & Controlled Y   \\ \hline
\texttt{U2} & 2 parameter 1 pulse 1-qubit  & \texttt{SWAP} & Swap gate   \\ \hline
\texttt{U1} & 1 parameter 0 pulse 1-qubit  &  \texttt{CH} & Controlled H   \\ \hline
\texttt{CX} & Controlled-NOT  &  \texttt{CCX} & Toffoli gate   \\ \hline
\texttt{ID} & Idle gate or identity  &  \texttt{CSWAP} & Fredkin gate \\ \hline
\texttt{X} & Pauli-X bit flip  & \texttt{CRX} & Controlled RX rotation  \\ \hline
\texttt{Y} & Pauli-Y bit and phase flip  & \texttt{CRY} & Controlled RY rotation  \\ \hline
\texttt{Z} & Pauli-Z phase flip  &  \texttt{CRZ} & Controlled RZ rotation \\ \hline
\texttt{H} & Hadamard  &  \texttt{CU1} & Controlled phase rotation  \\ \hline
\texttt{S} & sqrt(Z) phase  &  \texttt{CU3} & Controlled U3 \\ \hline
\texttt{SDG} & Conjugate of sqrt(Z)  &  \texttt{RXX} & 2-qubit XX rotation \\ \hline
\texttt{T} & sqrt(S) phase  & \texttt{RZZ} & 2-qubit ZZ rotation   \\ \hline
\texttt{TDG} & Conjugate of sqrt(S) & \texttt{RCCX} & Relative-phase CXX \\ \hline
\texttt{RX} & X-axis rotation & \texttt{RC3X} & Relative-phase 3-controlled X \\ \hline
\texttt{RY} & Y-axis rotation & \texttt{C3X} & 3-controlled X \\ \hline
\texttt{RZ} & Z-axis rotation & \texttt{C3XSQRTX} & 3-controlled sqrt(X) \\ \hline
\texttt{CZ} & Controlled phase & \texttt{C4X} & 4-controlled X \\ \hline
\end{tabular}
\label{tab:gates}
\end{table}

In terms of the frontend design, the state preparation algorithm described in Section~\ref{sec:algorithm} is implemented in Qiskit. Although SV-Sim directly supports Qiskit through the Python-API, given the extremely deep circuit, the parsing of the Qiskit gate object and performing gate conversion in Python can lead to considerable overhead. Consequently, we use the QASM frontend.  Table~\ref{tab:gates} lists the gates defined by QASM~\cite{cross2017open}. Given such gate support, and the gate fusion technique to be presented, when generating QASM, no transpilation optimization is needed in Qiskit, which also improves performance.


\subsection{Gate Fusion}
\label{subsec:gate_fusion}

\begin{figure}[!t]
\centering
\includegraphics[width=0.8\columnwidth]{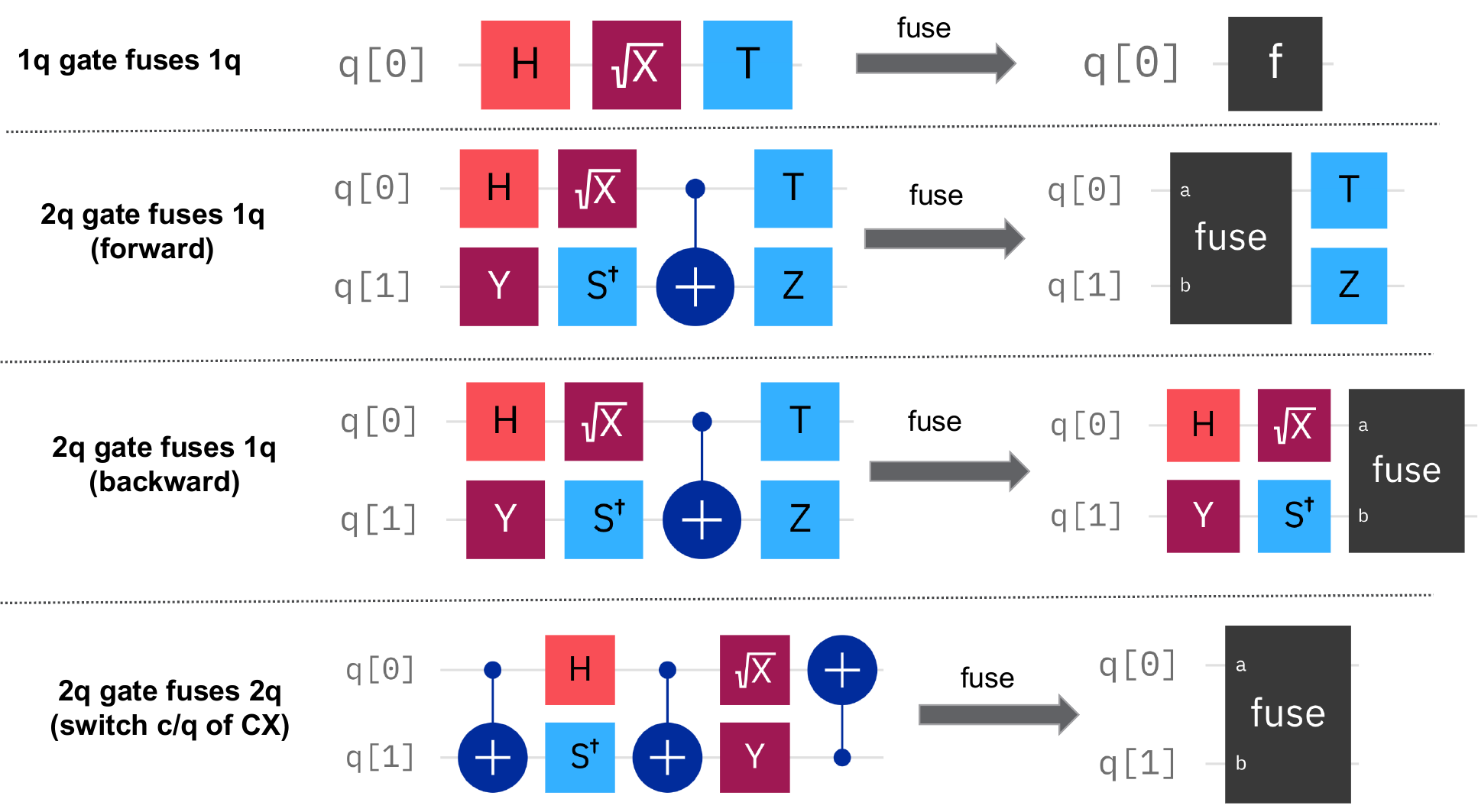} 
\caption{ 1-qubit and 2-qubit gate fusion operations for state-vector simulation.}
\label{fig:gate_fusion}
\end{figure}

The first challenge for the numerical simulation, as mentioned, is the deep circuit. We propose gate fusion to merge gates and shrink circuit depth. Given it is numerical simulations, we are not limited by the basis gates of a real device. We thus propose the following fusion operations:

\vspace{4pt}\noindent\emph{1-qubit gates fuse 1-qubit gates:} This implies that all consecutive 1-qubit gates applied on the same qubit can be merged into a single unitary gate (see Fig.~\ref{fig:gate_fusion}). We implement a general 1-qubit gate labeled as \texttt{C1} in the simulator to perform the merged unitary gate.

\vspace{4pt}\noindent\emph{2-qubit gates fuse 2-qubit gates:} Similarly, we can merge consecutive 2-qubit gates over the same qubit pairs together into a single 2-qubit unitary gate. This includes the conditions of switching the control and target qubit of \texttt{CX} gates, as shown in Fig.~\ref{fig:gate_fusion}. We realize a general 2-qubit gate labeled as \texttt{C2} in the simulator to run this gate.

\vspace{4pt}\noindent\emph{2-qubit gates fuse 1-qubit gates:} For state-vector simulation, it is also feasible for a 1-qubit gate to concatenate with a 1-qubit identity gate, forming a 2-qubit gate, and thus can be merged with another 2-qubit gate of the same qubit pair. This is similar to the 2-qubit gate ``absorbs" a 1-qubit gate. Depending on the order of the 1-and 2-qubit gate, a forward or a backward fusion can be established, see Fig.~\ref{fig:gate_fusion}. We use \texttt{C2} to execute the fused 2-qubit gate.

\begin{figure}[!t]
\centering
\includegraphics[width=0.9\columnwidth]{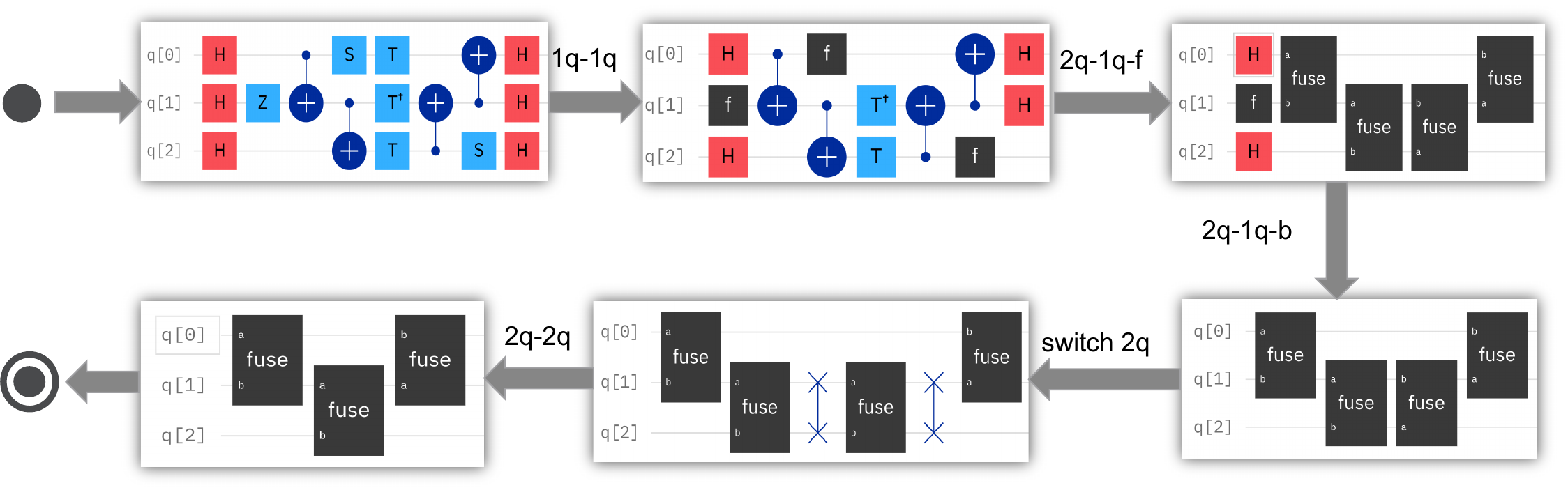} 
\caption{Proposed gate fusion strategy.}
\label{fig:gate_fusion_path}
\end{figure}

To effectively explore all fusion opportunities, we propose the following strategy to compose these fusion operations, shown in Fig.~\ref{fig:gate_fusion_path}. We start by first merging consecutive 1-qubit gates applying to the same qubit. Then, we conduct forward and backward absorption for the 2-qubit gates to fuse the surrounding 1-qubit gate. After that, we add \texttt{SWAP} gates to ensure all 2-qubit gates are having the same partial order that the first qubit index is smaller than the second, i.e., \texttt{U(b,a)}$\to$\texttt{SWAP(a,b)U(a,b)SWAP(a,b)}. This will facilitate 2-qubit 2-qubit fusion. We finally run 2-qubit fusion to obtain the ultimate circuit for simulation. We show in Section~\ref{sec:evaluation} about the efficiency of gate fusion.

\subsection{Mid-Circuit Measurement Assertion}
\label{subsec:mid_cir}

As discussed in Section~\ref{subsec:proj}, the projection algorithm works by asserting the ancilla qubit measuring $\ket{0}$, which will progressively enhance the amplitude of the ground state with respect to alternative states. When all ancilla qubits testing $\ket{0}$, the ground state is well prepared, which can be sampled. Listing~\ref{lst:qasm} shows the structure of the QASM circuit generated from the projection algorithm. $q[0]$ is the ancilla qubit, which is measured (Line~11) and recycled (Line~13) repeatedly in the middle of the circuit. After the execution, from the $m$ measurement results of $r$, the algorithm picks those (labeled $k$) with respect to $c$ being all-zero, which are the right samples of the ground state.

\begin{figure}[!t]
\begin{lstlisting}[caption=Block structure with repeated ancilla measurement and reset for the low-energy nuclear projection algoirthm in QASM,captionpos='b',label={lst:qasm}, basicstyle={\scriptsize\ttfamily\bfseries}, numberstyle=\scriptsize, backgroundcolor =\color{tinygray},framexleftmargin=10pt]
OPENQASM 2.0;
include "qelib1.inc";
qreg q[21];
creg c[11];
creg r[20];
//======= filtering step-1 =======
ry(7.36183164e-07) q[0];
ry(-7.579507071484729) q[0];
sdg q[0];
...//lots of gates
measure q[0] -> c[0];
barrier q;
reset q[0];
barrier q;
//======= filtering step-2 =======
ry(1.26927712e-06) q0[0];
ry(-7.593605206153606) q0[0];
sdg q0[0];
...//lots of gates
measure q[0] -> c[1];
...//filtering step-n
//======= Final measurement =======
measure q -> r; //sample the prepared state for m shots
\end{lstlisting}
\end{figure}

Looking at Listing~\ref{lst:qasm}, although the projection algorithm effectively reduces the ancilla usage to 1 qubit, which can significantly reduce simulation cost, it also introduces mid-circuit measurement and the potential dependency on the ancilla. For state-vector simulation of general circuits that only incurs measurement at the end, a well-known and key optimization is that one can simulate the circuit upon the measurement, and sample the final state-vector for the $m$ shots together at once. Here, because of the mid-circuit measurement and dependency, we can only run one shot a time, drastically degrading the simulation efficiency. If the probability of $c$ being all zero is relatively low (see Table~\ref{tab:mma}, we either cannot obtain the state of interest ($k=0$), or the sampling efficiency is low ($k\ll m$).

Therefore, in addition to gate fusion, we further modify SV-Sim to perform an algorithm-specific implementation. Recalling that for each ancilla measurement over $q[0]$, only $q[0]$ being $\ket{0}$ is of interest. Therefore, during simulation, after obtaining the probability of measuring $q[0]$, we can verify whether $q[0]$ still has any chances to be $\ket{0}$, i.e., $P(q[0]=\ket{0})\gt 0$. If so, we simply assert the measurement result of sampling $q[0]$ is state $\ket{0}$, and track its probability $P(q[0]=\ket{0})$. Otherwise, the simulation terminates and returns a failure. In this way, we can ensure $c$ is always all-zero at the end. More importantly, despite we still need mid-circuit measurement, the dependency over $q[0]$ is resolved. We can simply run the simulation to the end, and sample $m$ times at once, of which all of them are of interest (i.e., $m=k$), given $c$ is ensured to be zero. Through this design, we can significantly improve simulation efficiency as well as sampling efficiency for the projection filtering algorithm.

\section{Evaluation}
\label{sec:evaluation}


\subsection{Environment and Settings}

We use the U.S. NERSC Perlmutter HPC cluster and the OLCF Summit and Crusher HPC clusters for the evaluation, as listed in Table~\ref{tab:platform}.

\vspace{4pt}\noindent\textbf{Perlmutter:} We use the Perlmutter system as the primary platform for the evaluation. Perlmutter is an HPE Cray EX pre-exascale HPC system based on the HPE Cray Shasta platform. This heterogeneous cluster comprises 1536 GPU nodes (each contains 4 NVIDIA A100 40GB GPUs), 256 advanced GPU nodes (4 A100 80GB GPUs), and 3072 CPU nodes (AMD EPYC 7763) linked by HPE Slingshot 11 high-speed interconnect. Each node is equipped with an AMD Milan EPYC 7763 64-core CPU. 

\vspace{4pt}\noindent\textbf{Summit:} We also use the OLCF Summit cluster. This IBM AC922 system contains more than 27,000 NVIDIA Volta GPUs with more than 9,000 IBM POWER9 CPUs. It has 4,608 nodes. Each node features two IBM POWER9 CPUs with 512 GB DDR4 main memory, and 6 Volta V100 GPUs with 16GB HBM2 memory per GPU. The intra-node interconnect is NVLink and the inter-node network is EDR 100GB InfiniBand. 

\vspace{4pt}\noindent\textbf{Crusher:} Crusher is an early-access testbed system at ORNL for exascale computing. It shares identical hardware and similar software as the Frontier HPC. Crusher has 2 cabinets, with 128 compute nodes and 64 compute nodes, respectively. Each node consists of a 64-core AMD EPYC 7A53 CPU, 512 GB of DDR4 main memory, and 4 AMD MI250X GPUs. The CPUs and GPUs are connected through Infinity Fabric inside a node. Different nodes are connected through 4 HPE Slingshot 25 GB/s NICs.

\begin{table*}[!t]
\centering\footnotesize
\caption{Evaluation platforms: NERSC Perlmutter, OLCF Summit and OLCF Crusher HPC systems. IC refers to intra-node interconnect. Network refers to inter-node network. Runtime refers to GPU runtime.}
\begin{tabular}{|c|c|c|c|c|c|}
\hline
\textbf{Platform} & \textbf{Nodes} & \textbf{CPU} & \textbf{Cores} & \textbf{DRAM} & \textbf{Compiler}  \\ \hline
Perlmutter & 1500 & AMD EPYC-7763 & 64 & 256 GB & nvc++ 21.11  \\ \hline
Summit & 4608 & IBM Power-9 & 44 & 512 GB & gcc-9.3.0 \\ \hline
Crusher & 192 & AMD EPYC-7A53 & 64 & 512 GB & gcc-12.2.0 \\ \hline
\hline
\textbf{Platform} & \textbf{GPUs} & \textbf{GPU Memory} & \textbf{GPU IC} & \textbf{Network} & \textbf{Runtime} \\ \hline
Perlmutter & 4 NVIDIA A100 & 40 GB HBM2 &  NVLink-V3 & Slingshot  &  CUDA-11.5 \\ \hline
Summit & 6 NVIDIA V100 & 16 GB HBM2 & NVLink-V2 & InfiniBand & CUDA-11.1 \\ \hline
Crusher & 4 AMD MI250X & 64 GB HBM2 & InfinityFabric & Slingshot & ROCM-5.3.0 \\ \hline
\end{tabular}
\label{tab:platform}
\end{table*}

\vspace{4pt}\noindent Regarding the problem settings, for testing purposes in this investigation, we used the phenomenological interacting shell model. In this many-body framework one assumes that only a small number of valence nucleons are interacting in a restricted model space via a phenomenological interaction. The rest of the nucleons are assumed to constitute an inert core. To simplify even more the problem, we present in this paper test cases where only neutrons are active. Thus, we have considered here two neutrons in the $0p$ shell (six active single-particle states) using the  Cohen-Kurath interaction \cite{COHEN19651}, four neutrons in the  $1s0d$ model space (12 active single-particle states) using the ``universal sd"  Wildental interaction \cite{WILDENTHAL19845,PhysRevC.74.034315}, and four neutrons in the $1p0f$ shell (20 active states) using the modified KB3 interaction \cite{KB3,KB3-2}. Taking into account the $^4$He core for the $0p$ shell, the $^{16}$O core for the $1s0d$ shell, and the $^{40}$Ca core for the $1p0f$ shell, $^6$He, $^{20}$O, and $^{44}$Ca have been considered. For the $0p$ shell case, we also used a deformed Hartree-Fock solution computed with the code \texttt{SHERPA} \cite{PhysRevC.66.034301}.

\begin{table}[!t]
\centering\footnotesize
\caption{Problem circuits for low-energy nuclear ground state preparation.}
\begin{tabular}{c|c|c|c|c|c|c}
\hline\hline
Problem & Nucleus & Qubits & Trotter Steps & Filtering Steps & Gates & 2-Qubit Gates  \\ \hline
P1 & $^6$He & 7 &  8 & 4 & 11,939 & 7,088  \\ \hline
P2 & $^6$He & 7 &  14 & 4 & 20,885 & 12,404  \\ \hline
P3 & $^6$He & 7 &  26 & 4 & 38,777 & 23,036  \\ \hline
P4 & $^6$He & 7 &  47 & 4 & 70,088 & 41,642  \\ \hline
P5 & $^{20}$O & 13 & 14 & 8 & 210,457 & 126,980 \\ \hline
P6 & $^{20}$O & 13 & 24 & 8 & 360,767 & 217,680 \\ \hline
P7 & $^{20}$O & 13 & 44 & 8 & 661,387 & 399,080 \\ \hline
P8 & $^{20}$O & 13 & 83 & 8 & 1,247,596 & 752,810 \\ \hline

P9 & $^{44}$Ca & 21 & 18 & 8 & 1,970,931 & 1,208,052 \\ \hline
P10 & $^{44}$Ca & 21 & 30 & 8 & 3,284,871 & 2,013,420 \\ \hline
P11 & $^{44}$Ca & 21 & 58 & 8 & 6,350,731 & 3,892,612 \\ \hline
P12 & $^{44}$Ca & 21 & 108 & 8 & 11,825,481 & 7,248,312 \\ \hline
P13 & $^{44}$Ca & 21 & 214 & 8 & 23,431,951 & 14,362,396 \\ \hline
P14 & $^{44}$Ca & 21 & 528 & 8 & 57,813,381 & 35,436,192 \\ \hline
P15 & $^{44}$Ca & 21 & 1051 & 8 & 115,079,266 & 70,536,814 \\ \hline

\hline\hline
\end{tabular}
\label{tab:prsetting}
\end{table}

With that, we formalize 15 problem instances for evaluating the simulation performance, covering a wide range of problem size from 7-qubit (including 1 ancilla), 11,939 gates to 21-qubit (1 ancilla), 115,079,266 gates. Their features are listed in Table~\ref{tab:prsetting}. Their QASM file-size ranges from 272KB to 2.2GB, confirming that the circuit depth for these low-energy nuclear state preparation applications are much deeper than general circuits.  

Our evaluation covers the following aspects: (i) \emph{Gate Fusion}: We first evaluate the effectiveness of gate fusion, focusing on gate count reduction and simulation time reduction. (ii) \emph{Mid-Circuit Measurement Assertion}: We show the savings of measurement times, and the improvement on sampling efficiency and simulation performance. (iii) \emph{Performance across Platforms}: we show the simulation time on the CPUs and GPUs of the three HPC clusters. For the CPU performance, we show single-core single-thread, and multi-core with 4 threads using OpenMP. (iv) \emph{Ground Energy and Sampling Efficiency}: we show the obtained ground state energy of the 15 problem instances with respect to different trotter steps.


\subsection{Evaluation Results}

\begin{figure*}[!h]
     \centering
    \includegraphics[width=\textwidth]{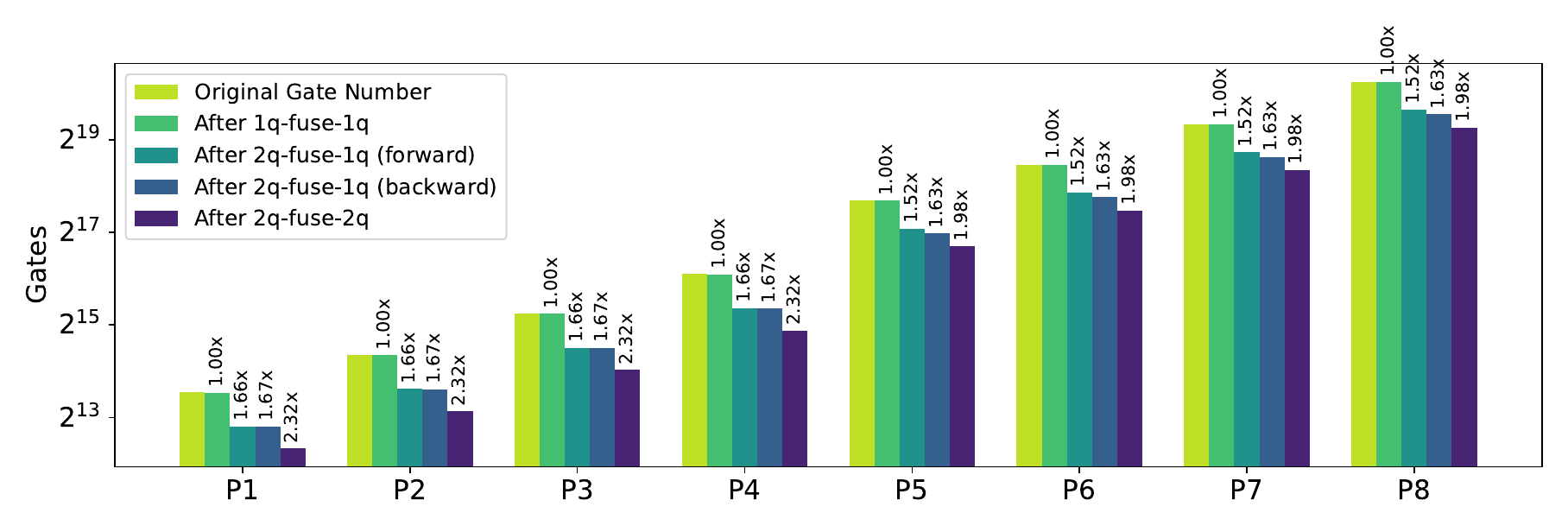}
         \caption{Reduction of gate count via the four gate fusion steps in Section~\ref{subsec:gate_fusion}.}
         \label{fig:gate_count_red}
\end{figure*}

\begin{figure*}[!h]
    \centering
    \includegraphics[width=\textwidth]{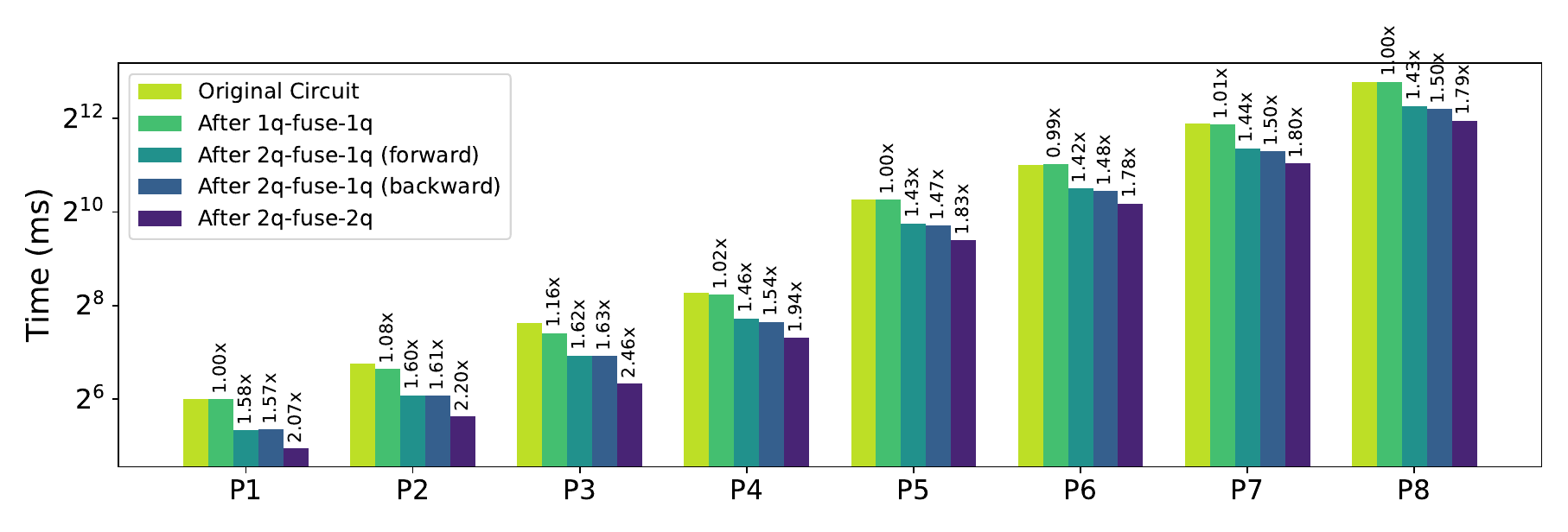}
    \caption{Performance improvement through the four gate fusion steps.}
    \label{fig:res_gate_fusion_sim}
\end{figure*}

\vspace{4pt}\noindent\textbf{(i) Gate Fusion:}
Using P1 to P8 in Table~\ref{tab:prsetting}, Fig.~\ref{fig:gate_count_red} illustrates the reduction of gate count through gate fusion presented in Section~\ref{subsec:gate_fusion}. We progressively apply each fusion operation of Fig.~\ref{fig:gate_fusion}, following the strategy in Fig.~\ref{fig:gate_fusion_path}. After that, Fig.~\ref{fig:res_gate_fusion_sim} shows the corresponding simulation time reduction with gate fusion, tested using an A100 GPU of Perlmutter.

As can be seen, gate fusion can bring on average $2.15\times$ gate count reduction (range: $1.98$-$2.32\times$) across the 8 problem instances, corresponding to an on-average $1.98\times$ speedup (range $1.78$-$2.46\times$). Regarding these results, we have the following observations: (i) Circuits for different nuclei have specific patterns which may largely impact the benefit from gate reduction. Trotter steps, on the other hand, do not seem to impact the benefit from gate reduction. (ii) For this projection filtering algorithm, the percentage of two-qubit gates is quite high (on average 60.53\%, see Table~\ref{tab:prsetting}). Comparatively, there are much more opportunities for \emph{2-qubit gates fuse 1-qubit gates} and \emph{2-qubit gates fuse 2-qubit gates} than \emph{1-qubit gates fuse 1-qubit gates}. The latter (2-qubit gates fuse 2-qubit gates) is particularly unique from general quantum circuits.

Overall, these results demonstrate that our gate fusion technique can effectively reduce the gate count and improve simulation performance for the low-energy nuclear ground state projection algorithm.

\vspace{4pt}\noindent\textbf{(ii) Mid-Circuit Measurement Assertion:} We use P9 as a non-trivial exemplar case for this evaluation. Shown in  Listing~\ref{lst:qasm}, the projection circuit includes a series of mid-circuit ancilla measurements to amplify the probability of the ground state, given the measurement result is 0. However, it is also possible that the measurement gives 1, which implies that the present shot fails to project to the correct state, seen as a rejection. As shown in Table~\ref{tab:prsetting}, the filtering steps for P9 is 8, implying 8 mid-circuit measurements, labeled as M1 to M8. In Table~\ref{tab:mma}, we repeat three simulations (Case-1 to 3) of the P5 circuit on a Perlmutter A100 GPU using 1024 shots without applying mid-circuit measurement assertion. The values under M1 to M8 show the number of rejected shots during each mid-circuit measurement. 

\begin{table}[!t]
\centering\footnotesize
\caption{Evaluation results for mid-circuit measurement assertion on P9.}
\begin{tabular}{c|c|c|c|c|c|c|c|c|c|c}
\hline\hline
 & M1 & M2 & M3 & M4 & M5 & M6 & M7 & M8 & Success & Time(s)  \\ \hline
Case-1 & 723 & 186 & 31 & 28 & 14 & 9 & 3 & 3 & 27  & 22553 \\ \hline
Case-2 & 701 & 166 & 54 & 21 & 30 & 11 & 0 & 0 & 41 & 24147 \\ \hline
Case-3 & 726 & 149 & 35 & 22 & 22 & 21 & 5 & 4 & 40 & 23891  \\ \hline

MMA & 0.29602 & 0.48617 & 0.69349 & 0.74823 & 0.73060 & 0.77238 & 0.93470 & 0.95811 &  0.037738 & 52.621 \\ 
\hline\hline
\end{tabular}
\label{tab:mma}
\end{table}

Within the three trails, for P9, on average 36 of the 1024 shots can project to the final state of interest, implying a sampling efficiency of 3.52\% for the projection algorithm. Additionally, because of repeated end-to-end runs due to mid-circuit measurement, the simulation time even on an A100 GPU is still quite significant, on-average 392 minutes. By applying mid-circuit measurement assertion, despite the probability of measuring all-zero remains low, we can always assert it measures 0 and tracking the probability. 

The last row of Table~\ref{tab:mma} shows the ground truth probability of measuring 0 during each mid-circuit measurement. Overall, the probability of obtaining all 0s for M1 to M8 is $3.77\%$. With mid-circuit measurement assertion, we can assert the probability, and ensure the ground state is well-prepared. We can then sample the state by 1024 shots at once in parallel, providing 1024 effective samples. Comparatively, to obtain the same number of effective samples, the original method would need $1024/0.0377=27,162$ shots on average and have to be executed sequentially due to mid-circuit measurement.

Overall, the simulation time with MMA is 52.621s, 447$\times$ faster than the baseline. When demanding the same number of effective samples, e.g., 1024, the speedup is about 12,719$\times$. Although the benefit is circuit dependent, it confirms the usefulness and effectiveness of mid-circuit measurement assertion. For larger problems or problems with even lower success rate, the gains can be even more extraordinary.

\begin{figure*}[!h]
     \centering
\includegraphics[width=\textwidth]{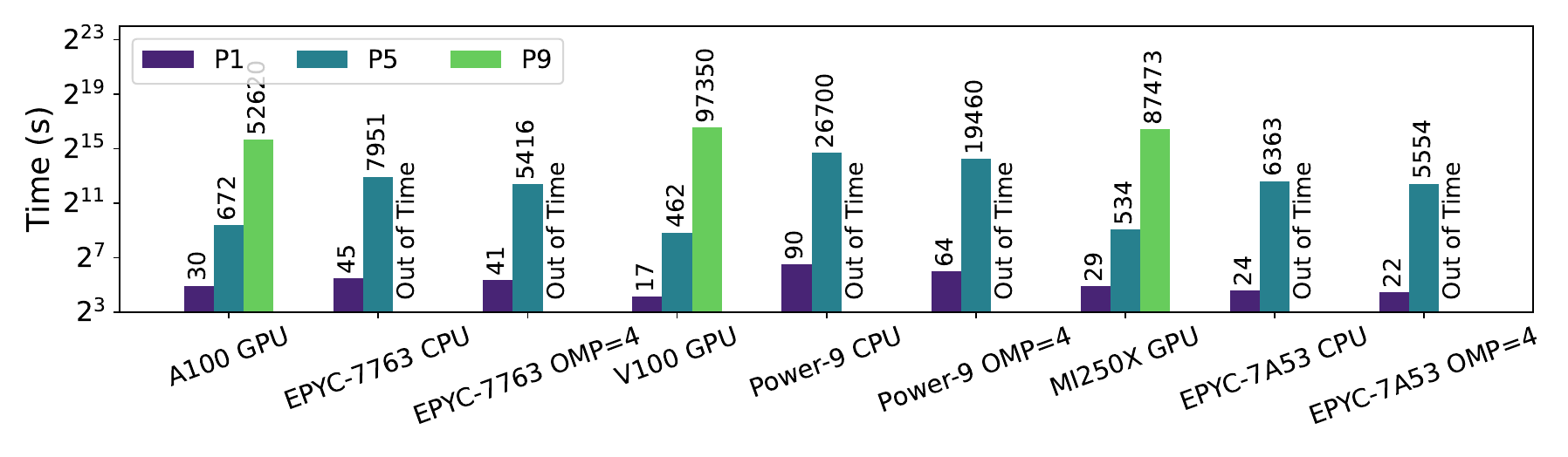}
         \caption{Simulation time for P1, P5 and P9 across the CPUs and GPUs of Perlmutter, Summit and Crusher.}
         \label{fig:cross_plat}
\end{figure*}

\noindent\textbf{(iii) Performance across Platforms:} We perform the simulations over the three CPUs and three GPUs of the Perlmutter, Summit and Crusher systems listed in Table~\ref{tab:platform}. Fig.~\ref{fig:cross_plat} shows the simulation time of the P1, P5 and P9 circuits on the three GPUs (A100, V100 and MI250X) and CPUs (EPYC-7763, Power-9 and EPYC-7A53 with 1 and 4 threads). We put the raw number of simulation latency (in ms) above the bars. Note, P9 runs too long on the CPUs so we omit them. As can be seen, for small cases such as 7-qubit P1, GPUs do not exhibit any performance advantages. For larger cases, such as 13-qubit P5, the performance of GPUs is well above CPUs. For even larger problems such as the 21-qubit P9, A100 is better than MI250X, and then V100, implying that A100 requires sufficient workload to deliver superior simulation performance.


\vspace{4pt}\noindent\textbf{(iv) Ground Energy and Sampling Efficiency:} Table~\ref{tab:energy} lists the ground state energy obtained through our simulation, the reference ground energy, and the theoretical successful rate of the projection filtering algorithm. We also list the GPU kernel execution time for the simulation. All the results here are obtained on an A100 GPU of Perlmutter.

\begin{table}[!t]
\centering\footnotesize
\caption{Results of ground state energy in MeV.}
\begin{tabular}{c|c|c|c|c|c|c}
\hline\hline
Problem & Nucleus & Trotter & Success Rate & Reference E(MeV) & Result E(MeV) & Kernel Time \\ \hline
P1 & $^6$He & 8 & 26.90\% & -3.90983149 &  -2.64541903 & 30.8ms \\ \hline
P2 & $^6$He & 14 & 40.87\% & -3.90983149 &  -3.36176356 & 39.4ms \\ \hline
P3 & $^6$He & 26 & 40.35\% & -3.90983149 &  -3.80798262 & 80.4ms \\ \hline
P4 & $^6$He & 47 & 37.33\% & -3.90983149 &  -3.90193879 & 158.4ms \\ \hline
P5 & $^{20}$O & 14 & 27.25\% & -35.26663264 &  -34.89780987 & 672.8ms \\ \hline
P6 & $^{20}$O & 24 & 38.17\% & -35.26663264 & -35.14916931 & 1.16s  \\ \hline
P7 & $^{20}$O & 44 & 41.69\% & -35.26663264 & -35.22824245 & 2.10s \\ \hline
P8 & $^{20}$O & 83 & 42.53\% & -35.26663264 &  -35.25485687 & 3.95s\\ \hline
P9 & $^{44}$Ca & 14 & 3.77\% & -5.00586901 & -4.33397239 & 52.62s \\ \hline
P10 & $^{44}$Ca & 30 & 10.01\% & -5.00586901 & -4.53559332 & 87.68s \\ \hline
P11 & $^{44}$Ca & 58 & 12.92\% & -5.00586901 & -4.85291636 & 169.50s \\ \hline
P12 & $^{44}$Ca & 108 & 13.60\% &-5.00586901 & -4.91713907 & 315.59s \\ \hline
P13 & $^{44}$Ca & 214 & 13.76\% &-5.00586901 & -4.93915187 & 625.60s \\ \hline
P14 & $^{44}$Ca & 528 & 13.76\% &-5.00586901 & -4.94418587 & 25.7min \\ \hline
P15 & $^{44}$Ca & 1051 & 13.74\% & -5.00586901 & -4.94454380 & 56.5min \\ \hline

\hline\hline
\end{tabular}
\label{tab:energy}
\end{table}

As can be seen, the high success rates indicate that the projection filtering algorithm is quite effective in converging to the ground state energy, even for non-trivial problems. Additionally, with only a few filtering steps, i.e., 4 for $^6$He and 8 for $^{20}$O and $^{44}$Ca, we can already achieve an accuracy with less than 0.01 MeV, 0.01 MeV, and 0.1 MeV, respectively. It also confirms that the simulator can generate the correct energy number with proper trotter steps. We also compare the SV-Sim energy results with Qiskit simulation results, they are aligned with each other to $10^{-14}$, further confirming the numerical accuracy of our simulation. Meanwhile, it is interesting to observe that for $^{44}$Ca, increasing the trotter steps from 528 to 1051 only brings very incremental gain of 0.0004 MeV, implying marginal impact. We set the investigation of impact from trotter steps as a future work.  



\section{Conclusion}
\label{sec:conclusion}

Numerical simulation represents an important method for direct verification of quantum circuit correctness. Here we have demonstrated verification of deep quantum circuits for nuclear physics applications using a high-performance numerical simulator that incorporates several unique methods. The first is the use of gate fusion to reduce the effective simulation depth and, therefore, reduce the amount of computation required to generate the quantum state. We have described and demonstrated the use of 1-qubit and 2-qubit gate fusion to reduce the simulation depth of circuits used to prepare nuclear state ansatz. The second method manages the simulated state in the presence of mid-circuit measurements. We simulated the role of mid-circuit measurement by following the complete simulated state and using post-selection of the designated ancilla to calculate the projected state as well as the probability of the outcome.  
\par 
Together these methods have permitted the verification of quantum circuits derived from a state preparation algorithm using an energy-filtering algorithm. We demonstrated simulation of several examples of deep circuits acting on 21 qubits with more than 115,000,000 gates. We have also tested these examples on a variety of high-performance computing systems that emphasize the benefits for GPU-accelerated processing. For future work, we would like to investigate the simulation of larger nucleus problems, such as $^{26}$Mg, $^{56}$Fe and $^{58}$Ni, etc., and the impact of accuracy from the filtering and the trotter steps.
\par
In conclusion, we have extended the capabilities for numerical simulation of deep quantum circuits for nuclear physics using high-performance computing systems. These methods provide powerful tools for verification of quantum circuit design.


\section*{Acknowledgement}
\label{sec:ack}
This material is based upon work supported by the U.S. Department of Energy, Office of Science, National Quantum Information Science Research Centers, Quantum Science Center (QSC). TSH, AL, and AB acknowledge QSC support for advances in numerical simulation methods and quantum circuit synthesis. The work of IS was carried out under the auspices of the National Nuclear Security Administration of the U.S. Department of Energy at Los Alamos National Laboratory under Contract No. 89233218CNA000001. IS gratefully acknowledge  partial support by the Advanced Simulation and Computing (ASC) Program.
This research used resources of the Oak Ridge Leadership Computing Facility (OLCF), which is a DOE Office of Science User Facility supported under Contract DE-AC05-00OR22725. This research used resources of the National Energy Research Scientific Computing Center (NERSC), a U.S. Department of Energy Office of Science User Facility located at Lawrence Berkeley National Laboratory, operated under Contract No. DE-AC02-05CH11231.

\appendix
\section{Implementation of the Ge {\it et al.} algorithm}
\label{app:cirac_algo}
We briefly review here the algorithm used for state preparation that follows the work of Ref.~\cite{Ge2018} and describe how has been implemented. We consider an Hamiltonian with spectrum in an interval $I\in(0,1)$, spectral gap $\Delta$, an initial trial state $\ket{\Phi}$ with overlap $\chi$ with the ground state $\ket{\Psi_0}$, and we assume we know the ground state energy $E$. After defining the shifted Hamiltonian $H^\prime=H-E\,\mathbf{1}$ the state defined below
\begin{eqnarray}
\ket{\tilde{\Phi}}=\frac{\cos^M(H^\prime)}{\lvert \lvert\cos^M(H^\prime) \rvert\rvert}\ket{\Phi} \label{eq:cosM-filter}
\end{eqnarray}
becomes close to the exact ground state, if the power of $M$ is chosen as in Ref.~\cite{Ge2018}. It is possible to approximate the operator in Eq. ~\eqref{eq:cosM-filter} as a linear combination of unitaries in the following way:
\begin{eqnarray}
\cos^{2m}H^\prime=\sum_{k=-m_0}^{m_0}\alpha_k e^{-2iHk}+R\, , \label{eq:cirac-expansion}
\end{eqnarray}
with 
\begin{eqnarray}
\alpha_k=\frac{1}{4^m}\binom{2m}{m+k}\, .
\end{eqnarray}
and 
$m_0$ properly chosen in order to vanishing of quantity $R$, a prescription is provided in Ref.~\cite{Ge2018}. The implementation of the above operator can be done using Linear Combination of Unitaries using the Prepare and Select oracles of Ref.~\cite{Roggero2020a} and briefly reviewed below. Given the $2m_0+1$ unitaries of Eq.~\ref{eq:cirac-expansion} we define an ancillary register of dimension $n_A=\lceil\log_2(2m_0+1)\rceil$. We can implement a block encoding of the linear combination of unitaries using the algorithm originally developed in Ref.~\cite{Childs2012,Childs2017}and using the approach reported in Ref.~\cite{Holmes2022}. In particular we first need a prepare operator $P$ acting only on the ancillary register defined by the following equation:
\begin{eqnarray}
P\ket{0}&=&\frac{1}{\sqrt{\alpha}}\sum_{k=0}^{2m_0}(\alpha_{-2m_0+k})^{1/2}\ket{k}\, ,
\end{eqnarray}
and $\alpha=\sum_{k=-m_0}^{m_0}\lvert\alpha_k\rvert$.
Following Ref.~\cite{Holmes2022} for the general case the gate decomposition of the unitary $P$  can be done using generic circuit synthesis
as originally reported in Ref.~\cite{Shende2006} whose exponential scaling should not be a limitation for the problems at hand (i. e. for $10^3$ unitaries  only $10$ ancillary qubits will be needed).  After we define the select oracle, acting both on the ancillary register and the target register, in the following way
\begin{eqnarray}
S&=&\sum_{k=0}^{2m_0+1}\ket{k}\bra{k}\otimes U^k\, ,
\end{eqnarray}
where $U=e^{-2iH}$. Although the implementetion in principle requires applying several multi-controlled unitaries, in Ref.~\cite{Holmes2022}, Lemma 3.5, has been shown that only a number of single controlled unitaries equal to the number of ancillary qubits are necessary and sufficient. Therefore the implementation of the LCU method for the problem at hand can be expressed as in Fig.~2 of Ref.~\cite{Holmes2022}. We are now in the position to discuss the success probability of the procedure similarly to Ref.~\cite{Roggero2020a} we can define the following quantity
\begin{eqnarray}
\eta^2&=&\bra{\Phi} O^2\ket{\Phi}\, ,
\end{eqnarray}
where we have defined the operator $O=\sum_{k=-m_0}^{m_0}\alpha_k U^k$. The success probability of postselecting all $0s$ on the ancillary qubit is 
\begin{eqnarray}
P_s&=&\frac{\eta^2}{\alpha^2}\, .
\end{eqnarray}




\bibliography{sn-bibliography}


\end{document}